\documentclass[journal]{IEEEtran}
\usepackage{cite}
\usepackage{graphicx}
\usepackage{color}
\usepackage{setspace} 
\usepackage[tight,footnotesize]{subfigure}
\usepackage{booktabs}
\usepackage{url}
\usepackage{tabularx}
\usepackage{tabularx}
\usepackage{xurl}

\ifCLASSINFOpdf
\else
\fi

\ifCLASSOPTIONcompsoc
  \usepackage[caption=false,font=normalsize,labelfont=sf,textfont=sf]{subfig}
\else
  \usepackage[caption=false,font=footnotesize]{subfig}
\fi

\usepackage[cmex10]{amsmath}
\usepackage{multirow, makecell}
\usepackage{multicol}
\usepackage{hhline}
\usepackage{array}
\begin{document}
\title{\LARGE{Flexible Training Workloads in Large-Scale AI Data Centers for
Transient-Stability Support in Transmission-Constrained Power Systems}}
\author{Jae-Kyeong~Kim,~\IEEEmembership{Member,~IEEE}%
\thanks{This work has been submitted to the IEEE for possible publication.}
\thanks{J.-K. Kim is with Korea Electrotechnology Research Institute, Uiwang, 16029, Korea (e-mail: jkkim@keri.re.kr).}%
}
\markboth{}%
{Shell \MakeLowercase{\textit{et al.}}: Bare Demo of IEEEtran.cls for Journals}
\maketitle

\begin{abstract}
The rapid expansion of large-scale artificial intelligence (AI) data centers is introducing substantial, concentrated, and rapidly varying loads into transmission-constrained power systems. Although such load variations are generally regarded as operational challenges, this paper presents an alternative perspective in which the upward load flexibility of AI data centers is utilized for transient-stability support. To this end, training-induced load increase (TILI) is proposed as a fast demand-side operating concept that initiates or resumes flexible AI training workloads following fault clearing to increase active-power demand at electrically effective locations. Using idealized aggregate load responses, the stabilizing mechanism of TILI is first clarified in a single-machine infinite-bus (SMIB) system, and its effectiveness is then evaluated in the IEEE 39-bus system and a large-scale Korean power system. Case study results across all three systems demonstrate that TILI can increase the transient-stability-constrained generation limit, with its effectiveness depending on response magnitude, activation delay, and electrical siting. Larger and earlier TILI responses at electrically effective buses are more effective in improving transient stability. These results suggest that the upward load flexibility of AI data centers could provide complementary transient-stability support when sufficient electrical headroom, flexible training workloads, and reliable activation are available.
\end{abstract}

\IEEEpeerreviewmaketitle

\section{Introduction}
The rapid expansion of artificial intelligence (AI), particularly large language models and other foundation models, is creating a major new source of electricity demand. Training and inference for these models increasingly rely on large, high-density computing clusters equipped with graphics processing units
\cite{LLM_few_shot,LLM_compute},
accelerating the global deployment of large-scale AI data centers
\cite{IEA_2025}.
Individual facilities may require several hundred megawatts or more
\cite{IEA_2025,NERC_DS},
and their geographic concentration can intensify transmission congestion and impose additional constraints on power system planning and operations
\cite{EPRI_data_center,EU_Grid_DS}.
Furthermore, unlike conventional data center loads, which are often treated as relatively steady \cite{past_ds},
AI data-center demand can vary substantially with the composition and scheduling of training and inference workloads
\cite{LLM_charc,LLM_cloud}.
These workload-dependent variations can produce rapid changes in power demand over short timescales, creating new challenges for power balance, voltage regulation, and system stability~\cite{LLM_unseen,ds_osc}.

The impact of large-scale AI data centers on power systems depends not only on their power demand but also on their electrical location relative to generation resources and transmission constraints
\cite{NERC_Risk}.
In many power systems, major generation resources are concentrated far from large load centers, requiring substantial interregional power transfers through transmission corridors subject to thermal, voltage, and stability limits
\cite{system_CN,system_AU,system_UK,DOE,IEA_2020, KPS_ESS}.
Under these transfer-dependent operating conditions, a severe disturbance, such as a transmission-line fault, can reduce transfer capability and cause generators in a power-exporting region to accelerate and lose synchronism~\cite{kundur_stab}.
To prevent such instability, aggregate generation in the exporting region may need to be maintained below its installed capacity even when the generating units remain available
\cite{KPS_ESS}.
Large-scale AI data centers located electrically remote from such generation-rich regions can further increase transmission loading and constrain generation deliverability.

\begin{figure*}
\centering
\subfigure[AI data-center load response to training-workload initiation]{\includegraphics[width=2.8in]
{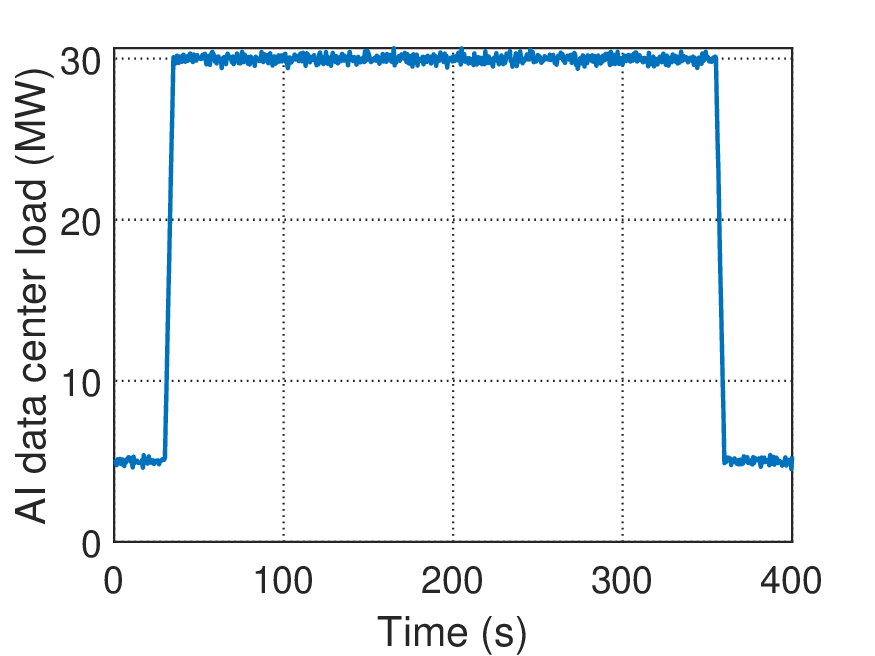}}
\subfigure[Operating principle of TILI for transient-stability support]{\includegraphics[width=3.8in]
{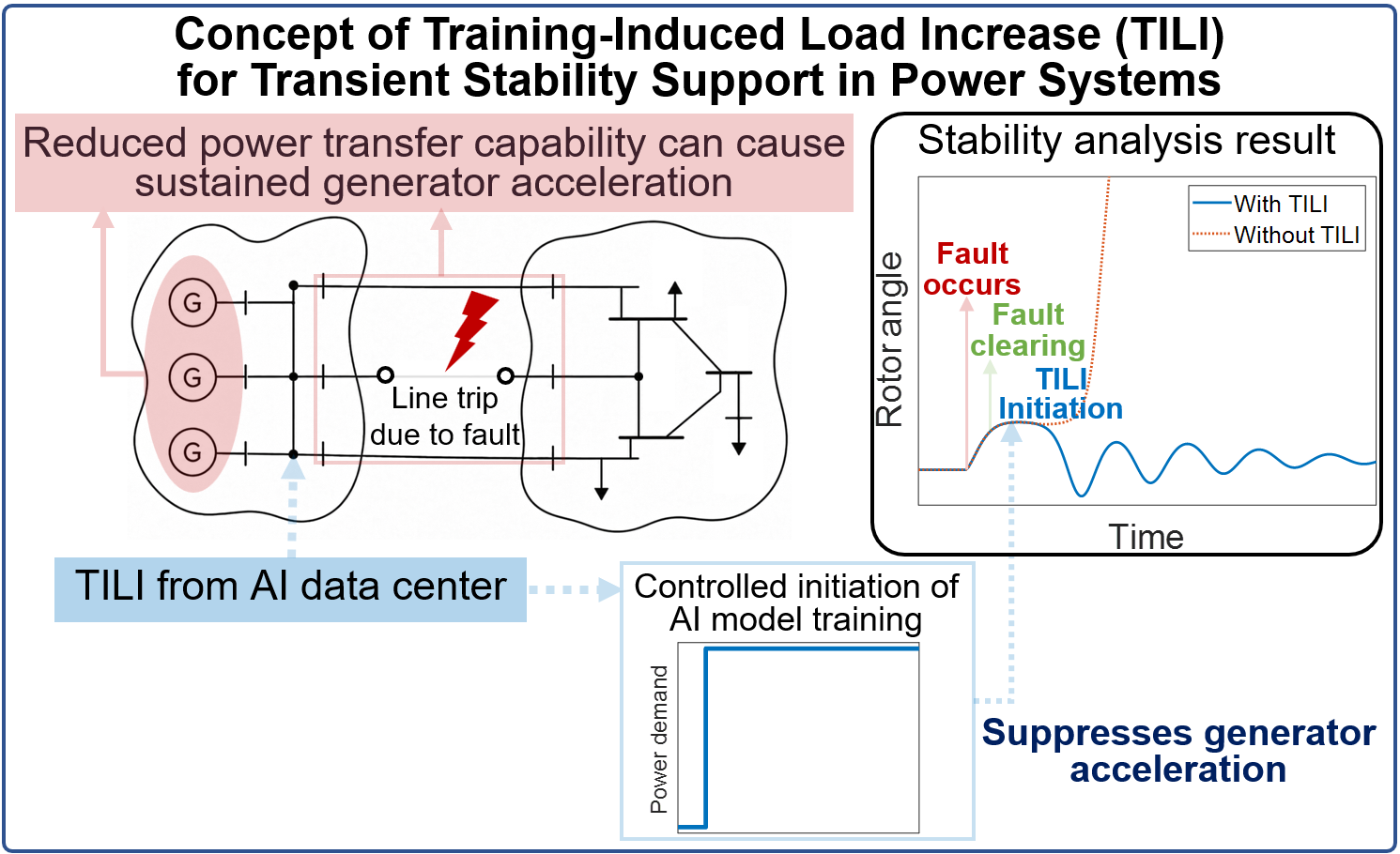}}
\caption{Measured AI data-center load response to training-workload initiation and operating principle of TILI for post-fault transient-stability support: (a) Measured active-power demand during the initiation of AI model training at a 50 MW AI data center in the United States \cite{NERC_DS} and (b) Conceptual mechanism of TILI for limiting post-fault generator acceleration and maintaining synchronism.}\label{fig00}
\end{figure*}

Therefore, AI data centers have generally been viewed as additional operational challenges for modern power systems. Their large, concentrated demand can increase network loading, while rapid workload-dependent power variations can complicate reliable system operation \cite{NERC_DS,LLM_unseen,ds_osc}.
The measured power trace shown in Fig.~\ref{fig00}(a) provides a representative example of how initiating large-scale AI training workloads can produce substantial increases in active-power demand over short timescales
\cite{NERC_DS}.
Existing studies on grid-interactive data centers have primarily focused on mitigating these effects through load smoothing, temporal or spatial workload shifting, and temporary demand curtailment
\cite{ai_grid_asset,ai_power_stab,LLM_cloud}.
These approaches mainly use demand flexibility to reduce, delay, or reshape electricity consumption. By contrast, the use of \emph{upward demand flexibility} as a fast transient-stability resource has received limited attention.

This paper investigates a complementary perspective in which the rapid increase in large-scale AI data-center demand, typically considered an operational challenge, is deliberately coordinated to support post-disturbance transient stability of power systems. Following fault clearing, synchronous-generator acceleration is governed by the imbalance between mechanical power input and electrical power output. When the post-fault transmission network cannot transfer sufficient electrical power, generators in the exporting region may undergo excessive first-swing acceleration. Rapidly increasing demand at electrically effective locations allows these generators to supply additional electrical power, thereby reducing the accelerating-power imbalance and limiting the first-swing rotor-angle excursion. The observed rapid power ramps of AI training workloads (e.g., Fig.~\ref{fig00}(a)) suggest that flexible computing workloads can be adjusted on timescales relevant to post-fault transient stability. Motivated by this capability, this paper investigates whether controlled load increases at AI data centers can improve transient stability, and how the resulting benefit depends on response magnitude, activation delay, and electrical siting.

To this end, this paper proposes training-induced load increase (TILI), a fast demand-side operating concept in which flexible AI training workloads are initiated or resumed following fault clearing. Fig.~\ref{fig00}(b) illustrates the overall concept and operating sequence of TILI. During the critical post-fault period, TILI rapidly increases AI data-center active-power demand at electrically effective locations. Unlike conventional demand-response strategies that primarily reduce or shift electricity consumption, the proposed TILI repurposes upward load flexibility as a fast corrective resource, increasing local demand and thereby the electrical output of accelerating generators. TILI is intended to complement established stability-enhancing measures, and its effectiveness is governed by response magnitude, activation delay, and electrical siting.

The proposed strategy is evaluated using three power systems of increasing complexity. First, a single-machine infinite-bus (SMIB) system is used to clarify the underlying physical mechanism and quantify the effect of TILI on the transient-stability-constrained generation limit. The analysis is then extended to the IEEE 39-bus system to examine the effectiveness of TILI in a multi-machine network and evaluate the effects of activation delay and electrical siting. Finally, TILI is assessed using a large-scale Korean power system in which regional generation is operationally constrained to maintain transient stability
\cite{KPS_ESS}.
Results from all case studies demonstrate that TILI can increase the transient-stability-constrained generation limit. Larger TILI responses, earlier activation, and siting at buses with a stronger electrical
influence on the critical generators provide greater generation-limit increases, whereas delayed activation and electrically remote siting reduce the stabilizing effect. To the best of the author's knowledge, the deliberate use of upward load flexibility from AI training workloads to support transient stability has not been systematically investigated. Therefore, the main contributions of this paper are summarized as follows:
\begin{itemize}
\item This paper reframes rapid, controllable load increases at AI data centers, typically regarded as operational challenges, as a potential resource for transient-stability support.
\item TILI is proposed as a grid-triggered demand-side strategy that initiates or resumes flexible AI training workloads following fault clearing.
\item The effectiveness of TILI is evaluated using the SMIB, IEEE 39-bus, and large-scale Korean power systems, quantifying how response magnitude, activation delay, and electrical siting affect transient-stability improvement.
\item Practical deployment requirements and remaining validation needs for TILI are identified, including electrical headroom, pre-staged training workloads, and reliable grid-triggered activation.
\end{itemize}

The remainder of this paper is organized as follows. Section~\ref{method} reviews the relevant transient-stability fundamentals and introduces the proposed TILI strategy. Section~\ref{casestudy} presents case-study results for the SMIB, IEEE 39-bus, and Korean power systems. Section~\ref{discussion} discusses the implications of TILI for power-system stability support, along with its deployment requirements, validation needs, and applicability. Section~\ref{conclusion} concludes the paper.

\section{TILI Principle and Modeling}\label{method}
\subsection{Transient-Stability Constraint Mechanism}
\begin{figure}
\centering
\subfigure[SMIB system with sending-end TILI]{\includegraphics[width=2.6in]{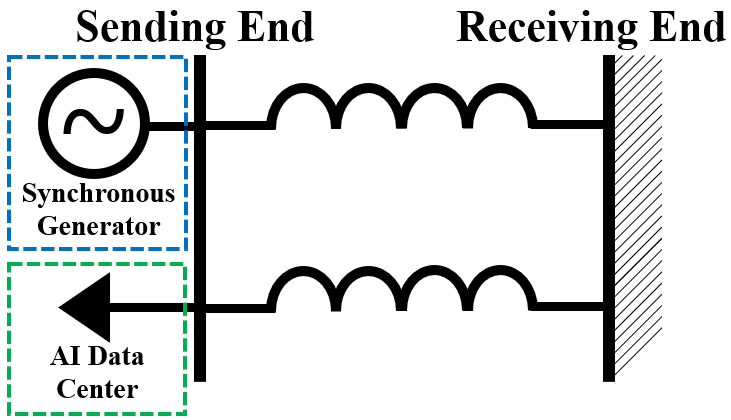}}
\subfigure[Rotor-angle responses]{\includegraphics[width=2.6in]{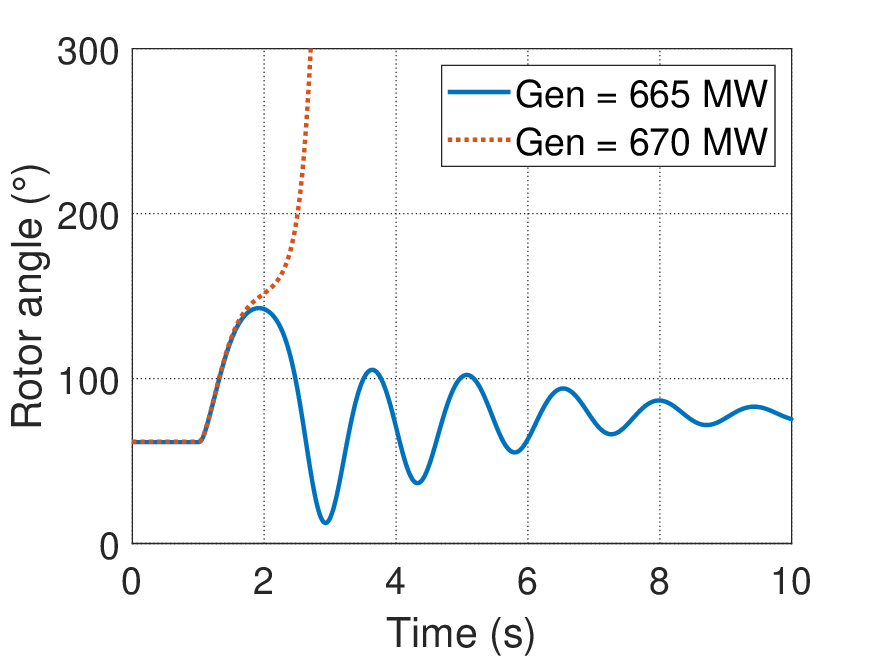}}
\caption{Schematic of the SMIB system and rotor-angle responses under the specified contingency: (a) system configuration with TILI at the sending end and (b) rotor-angle responses at generator outputs of 665 and 670~MW.}
\label{fig01}
\end{figure}

The transient-stability constraint considered in this study is illustrated using the SMIB system shown in Fig.~\ref{fig01}(a). A synchronous generator and a controllable local load representing an AI data center are connected at the sending end, whereas the receiving end is represented by an infinite bus. The transmission corridor initially consists of two parallel lines. The generator electrical output is divided into the power transferred through the transmission corridor and the power consumed locally at the sending end. Under the classical generator representation and a lossless network approximation, the rotor dynamics and active-power balance are expressed as follows:
\begin{align}
P_m - P_e
&= \frac{2H}{\omega_s}\frac{d^2\delta}{dt^2},
\label{eq01}\\
P_e
&= P_{\mathrm{trans}} + P_{\mathrm{local}},
\label{eq02}\\
P_{\mathrm{trans}}
&= \frac{E'V_R}{X_{\mathrm{eq}}}\sin\delta,
\label{eq03}
\end{align}
where $P_m$ and $P_e$ denote the generator mechanical input and electrical output powers, respectively. $H$ is the inertia constant; $\omega_s$ is the synchronous angular frequency; and $\delta$ is the generator rotor angle relative to the infinite bus. The variables $E'$ and $V_R$ denote the generator internal voltage and the receiving-end voltage, respectively, and $X_{\mathrm{eq}}$ is the equivalent transfer reactance between the generator internal voltage
and the receiving-end bus. The terms $P_{\mathrm{trans}}$ and $P_{\mathrm{local}}$ represent the active power exported through the transmission corridor and the active power consumed locally at the sending end, respectively. Under normal operating conditions, the generator operates close to steady state, such that the mechanical input power approximately balances the electrical output power, i.e., $P_m \approx P_e$. Therefore, the accelerating power in \eqref{eq01} is nearly zero, and the rotor speed and angle remain bounded around their operating values. 

However, when a fault occurs in the SMIB system, the associated voltage depression reduces the power transferred through the transmission corridor, $P_{\mathrm{trans}}$, and consequently decreases the generator electrical output $P_e$. Because the mechanical input $P_m$ changes negligibly over the first-swing timescale, the resulting imbalance $P_m-P_e$ becomes positive. The generator therefore accelerates, causing its rotor angle to increase. If the fault is cleared by tripping one of the two transmission lines, the post-fault network is weaker than the pre-fault network because the equivalent transfer reactance $X_{\mathrm{eq}}$ increases. Consequently, the maximum power that can be transferred through the remaining transmission corridor is reduced even after fault clearing. If the post-fault electrical output remains below the mechanical input, positive accelerating power persists, thereby causing the rotor angle to continue increasing. Transient instability occurs when the post-fault system cannot provide sufficient decelerating power to limit the rotor-angle excursion, eventually causing the generator to lose synchronism with the grid.

As an illustrative example, consider the SMIB system shown in Fig.~\ref{fig01}(a), in which the synchronous generator is rated at 1,241~MVA and the impedance of each parallel transmission line is 0.1~p.u. The base contingency consists of a three-phase fault at the sending-end bus, cleared after 0.1~s by tripping one of the two transmission lines. Fig.~\ref{fig01}(b) compares the generator rotor-angle responses for pre-fault active-power outputs of 665 and 670~MW. At 665~MW, the rotor angle remains bounded following fault clearing, indicating that the generator maintains synchronism after the contingency. In contrast, at 670~MW, the rotor angle diverges, indicating loss of synchronism and transient instability. These results show that the generator output is limited by the transient-stability constraint rather than by its modeled apparent-power rating. Therefore, within the 5~MW search resolution used in this study, 665~MW is adopted as the baseline transient-stability-constrained generation limit for the modeled SMIB system.

Maintaining transient stability therefore requires limiting the positive rotor acceleration during the critical first-swing interval following fault clearing
\cite{kundur_stab}.
As indicated by \eqref{eq01}, this requires the mismatch between the mechanical input power $P_m$ and the electrical output power $P_e$ to be reduced sufficiently rapidly. From \eqref{eq02}, a rapid increase in $P_{\mathrm{local}}$ provides a demand-side means of reducing this mismatch. This observation provides the physical basis for the proposed TILI.

\subsection{Operating Principle of the Proposed TILI}

Large-scale AI training workloads are typically managed by software-based workload schedulers. 
Subject to available computing capacity and electrical headroom, pre-staged or paused training jobs can be initiated or resumed on command. This controllability allows an AI data center to provide a rapid upward load response following a grid disturbance. The proposed strategy exploits this capability by initiating or resuming flexible AI training workloads after fault clearing. 

The resulting increase in AI data-center demand raises the local power-consumption term $P_{\mathrm{local}}$ in \eqref{eq02}, thereby increasing the generator electrical output $P_e$ during the critical post-fault interval. When TILI is activated at an AI data center electrically close to the affected generator, a larger portion of the additional demand is supplied by the affected generator without requiring an equivalent increase in power transfer through the weakened transmission corridor. The resulting increase in $P_e$ reduces the accelerating power $P_m-P_e$ in \eqref{eq01}, thereby mitigating rotor acceleration and limiting the first-swing rotor-angle excursion.

Through this mechanism, TILI repurposes rapid increases in large-scale AI data-center demand, which are generally regarded as a potential grid-reliability challenge \cite{LLM_unseen,ds_osc,NERC_DS}, as a targeted demand-side response for post-disturbance transient-stability support.

\subsection{Electrical Siting of TILI}
The stabilizing effect of TILI depends strongly on the electrical location of the participating AI data center. TILI is most effective at locations where the induced load response most strongly raises the electrical output of generators undergoing post-fault acceleration. Therefore, siting should be based on the electrical influence of the additional demand on the stability-constrained generators, rather than geographic proximity alone.

This location dependence follows from \eqref{eq01}--\eqref{eq03}. In the SMIB system, a controllable load connected at the sending end contributes directly to the local power-consumption term $P_{\mathrm{local}}$. An increase in $P_{\mathrm{local}}$ raises the generator electrical output $P_e$ without requiring an equivalent increase in power transfer through the weakened transmission corridor. Consequently, the accelerating power $P_m-P_e$ in \eqref{eq01} is reduced. By contrast, a load increase at the receiving end does not directly increase the local demand supplied by the sending-end generator under the SMIB representation. Because the power transferred through the corridor remains constrained by the post-fault equivalent reactance $X_{\mathrm{eq}}$, receiving-end TILI does not appreciably increase the electrical output of the affected generator and therefore provides essentially no first-swing stabilization.

Therefore, large-scale AI data centers intended to provide TILI should be sited at electrically effective buses where the induced load increase can be supplied predominantly by stability-constrained generators undergoing post-fault acceleration. In the SMIB case study, the AI data center is therefore located at the sending-end bus, as shown in Fig.~\ref{fig01}(a), so that the TILI response directly increases local power absorption and the electrical output of the affected generator.

\subsection{TILI Modeling and Activation Delay}
In this study, TILI is modeled as an idealized step increase in the aggregate active-power demand of the AI data center, applied after a prescribed activation delay following fault clearing. The activation delay represents the effective end-to-end response time from fault clearing to the establishment of the commanded load increase, consequently encompassing disturbance detection, signal transmission, workload scheduling, and load activation. The step magnitude represents the aggregate increase in demand assumed to be available from flexible AI training workloads. The response magnitude and activation delay are varied systematically to quantify how different TILI response characteristics affect transient-stability improvement.

It is worth noting that this representation isolates the effects of TILI response magnitude and activation timing without explicitly modeling finite load ramp-up, staged server activation, or workload-level dynamics. In the case studies, selected activation delays serve as baselines, and additional delays are used to quantify the sensitivity of TILI performance to response timing. Earlier activation enables the additional demand to reduce the accelerating-power imbalance at an earlier stage of the critical post-fault period. Therefore, the case study results characterize the system-level stability benefit associated with the assumed aggregate load response rather than the detailed dynamics of workload activation. 
Whether a particular AI data center can deliver the specified response magnitude within the prescribed activation delay requires separate validation under practical operating conditions.

\subsection{Stability Assessment and Evaluation Metrics}
Transient stability is assessed from the post-fault rotor-angle trajectories obtained through time-domain simulations. A case is classified as stable when the rotor-angle trajectories remain bounded after fault clearing and as unstable when one or more generators lose synchronism. The stabilizing effect of TILI is quantified by the increase in the maximum pre-fault generation that can be accommodated without loss of synchronism under the specified contingency. Accordingly, the post-fault rotor-angle trajectories are used as the primary stability indicator.

\section{Case Study Results}\label{casestudy}
\subsection{Case Study Overview}
Three power systems are considered to evaluate the stabilizing effect of TILI, the influence of AI data-center siting, and the sensitivity to activation delay: the SMIB system, the IEEE 39-bus system, and a large-scale Korean power system.
\subsubsection{Contingency Scenarios}
\begin{itemize}
    \item \textit{SMIB system:}
    A three-phase fault is applied at the sending-end bus and cleared
    after 0.1~s by tripping one of the two parallel transmission lines.
    \item \textit{39-bus system:}
    A three-phase fault is applied at Bus~10 and cleared after 0.1~s by
    tripping the transmission line between Buses~10 and~11.
    \item \textit{Korean system:}
    A severe 765-kV transmission-line contingency used in current
    operational security assessments of the Korean power system is
    simulated. This contingency imposes a transient-stability limit on
    generation in the eastern region.
\end{itemize}

\subsubsection{TILI Activation Delays}
In all case studies, TILI follows the idealized step-response model described in Section~\ref{method}-D. The system-specific baseline activation delays are defined as follows:
\begin{itemize}
    \item \textit{SMIB and 39-bus systems:}
	The baseline activation delay after fault clearing is set to 0.05~s. Note that 0.05-s baseline represents a fast-response reference case, and additional activation delays are considered to evaluate the effect of response timing on TILI performance.
	\item \textit{Korean system:}
	The baseline activation delay is set to four cycles
($\approx 0.0667$~s) after fault clearing, consistent with the response time of the special protection scheme (SPS) for generator tripping used in Korean system operation.
\end{itemize}

Additional activation delays are considered in the delay-sensitivity analyses to evaluate the effect of response timing on TILI performance.
\subsubsection{TILI Siting Configurations}
\begin{itemize}
    \item \textit{SMIB system:}
    The AI data center is connected at the sending-end bus, as shown in Fig.~\ref{fig01}(a), so that the induced load increase directly raises the electrical output of the stability-constrained generator.
    \item \textit{39-bus system:}
    Two electrically distinct TILI locations are considered. A 100 MW TILI response is alternatively applied at Bus~32, the terminal bus of the stability-constrained generator, and at Bus~10, an electrically more remote location. The two siting cases are evaluated under identical operating conditions, contingency scenarios, and TILI response magnitudes to isolate the effect of electrical siting.

\item \textit{Korean system:}
    Two siting configurations are considered within the eastern generation region. In the distributed configuration, seven existing loads with a total demand of approximately 653 MW are modeled as AI data centers. In the near-generator configuration, the same aggregate TILI responses are deployed at buses electrically closer to the generators that dominate the post-fault instability. Aggregate TILI magnitudes of approximately 65, 130, 196, 261, 326, 392, 457, and 522~MW are evaluated, corresponding to 10\%--80\% of the total modeled AI data-center demand.
\end{itemize}

\subsubsection{Stability Criteria and Performance Metrics}
The stability assessment follows the criteria defined in Section~\ref{method}-E, with time-domain simulations performed using PSS/E
\cite{PSSE}.
The system-specific performance metrics are defined as follows:
\begin{itemize}
    \item \textit{SMIB system:}
    The stability-constrained generation limit is defined as the maximum sending-end generation level at which the generator remains synchronized following the specified contingency. This limit is determined both without and with TILI, and the resulting increase in allowable generation is used to quantify the stabilizing effect of TILI. For selected target generation levels, the TILI magnitude required to maintain stability is also evaluated under different activation delays.
    \item \textit{39-bus system:}
    The stability-constrained generation limit is defined as the maximum output of the generator at Bus~32 that can be accommodated without loss of synchronism. This limit is re-evaluated with a 100 MW TILI response applied separately at each candidate location. The increase relative to the no-TILI case is used to quantify the stabilizing effect of TILI and compare the effectiveness of the two siting configurations.
    \item \textit{Korean system:}
    The eastern-area generation limit is defined as the maximum aggregate generation in the region for which synchronism is maintained following the specified 765-kV contingency. Aggregate generation in the eastern region is increased in discrete steps, while generation outside the region is reduced in accordance with the current Korean operating rule.

System stability is monitored using the systemwide rotor-angle spread, defined as the difference between the maximum and minimum generator rotor angles. A bounded spread indicates that the generators remain synchronized, whereas a diverging spread indicates loss of synchronism. The effectiveness of TILI is quantified by the resulting increase in the eastern-area generation limit across the considered TILI magnitudes, siting configurations, and activation delays.
\end{itemize}

\subsection{SMIB System}
The proposed TILI strategy is first examined using the SMIB system shown in Fig.~\ref{fig01}(a), which provides a transparent setting for illustrating its grid-stabilizing mechanism. Following a severe fault cleared by tripping one of the two transmission lines, the reduced post-fault transfer capability may cause the generator to accelerate and ultimately lose synchronism. Based on the rotor-angle responses in Fig.~\ref{fig01}(b), the baseline transient-stability-constrained generation limit is set to 665~MW. This benchmark is used to evaluate how TILI suppresses generator acceleration and increases the allowable generation.

\begin{figure}
\centering
\subfigure[TILI load response]{\includegraphics[width=2.6in]{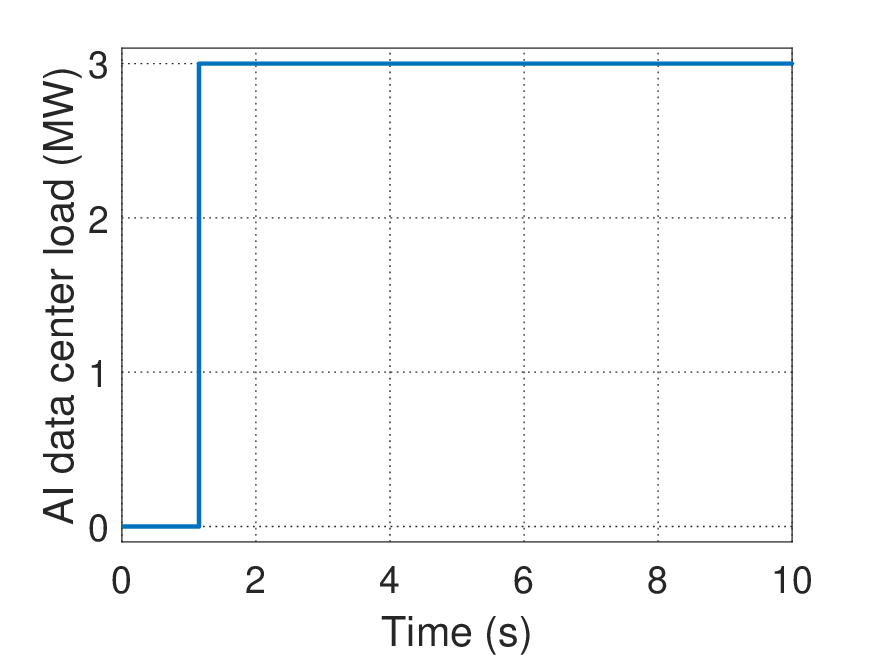}}
\subfigure[Rotor angle response]{ \includegraphics[width=2.6in]{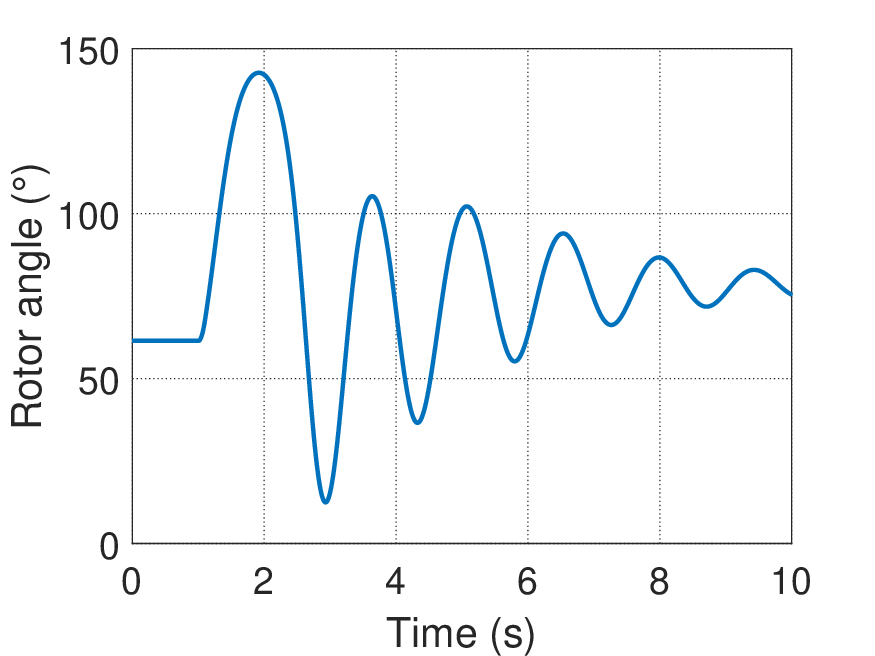}}
\caption{Stabilizing effect of a 3 MW TILI response in the SMIB system at a generator output of 670 MW: (a) AI data-center load response and (b) sending-end generator rotor-angle response.}\label{fig02}
\end{figure}

Fig.~\ref{fig02} illustrates how a rapid increase in AI training demand can stabilize an operating point above the baseline transient-stability-constrained generation limit. As shown in Fig.~\ref{fig02}(a), TILI is activated at the sending-end bus 0.05~s after fault clearing to provide a 3~MW load increase. The 670~MW operating point is unstable without TILI, as observed in Fig.~\ref{fig01}(b). Under the same contingency, however, Fig.~\ref{fig02}(b) shows that the generator remains synchronized when TILI is activated. The resulting increase in local demand raises the generator electrical output during the first swing, reduces the accelerating-power imbalance, and thereby stabilizes the otherwise unstable operating point.

\begin{figure}
\centering
\subfigure[Rotor-angle responses for different TILI activation delays]{
    \includegraphics[width=2.6in]{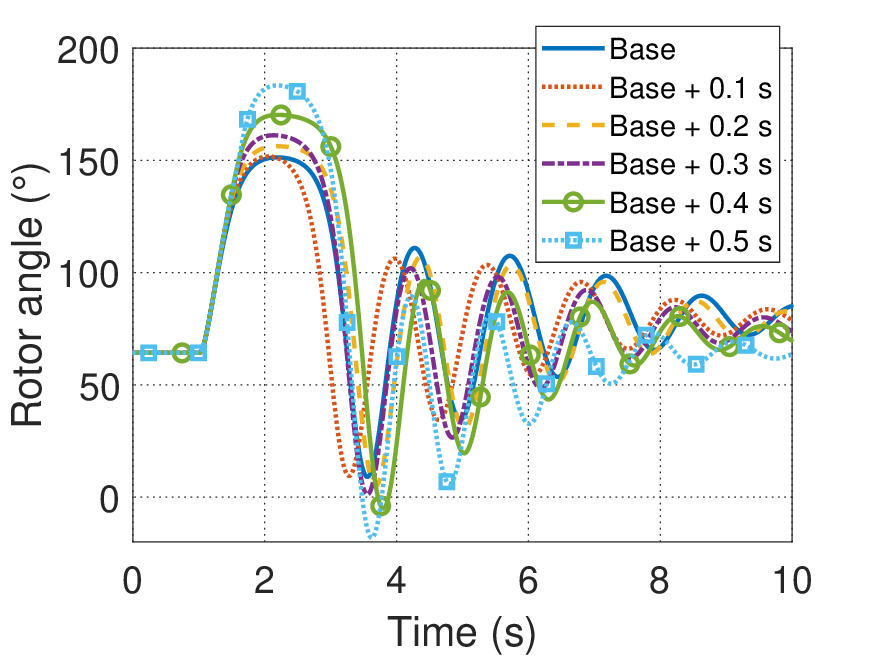}}
\subfigure[TILI magnitude required for stabilization]{
    \includegraphics[width=2.6in]{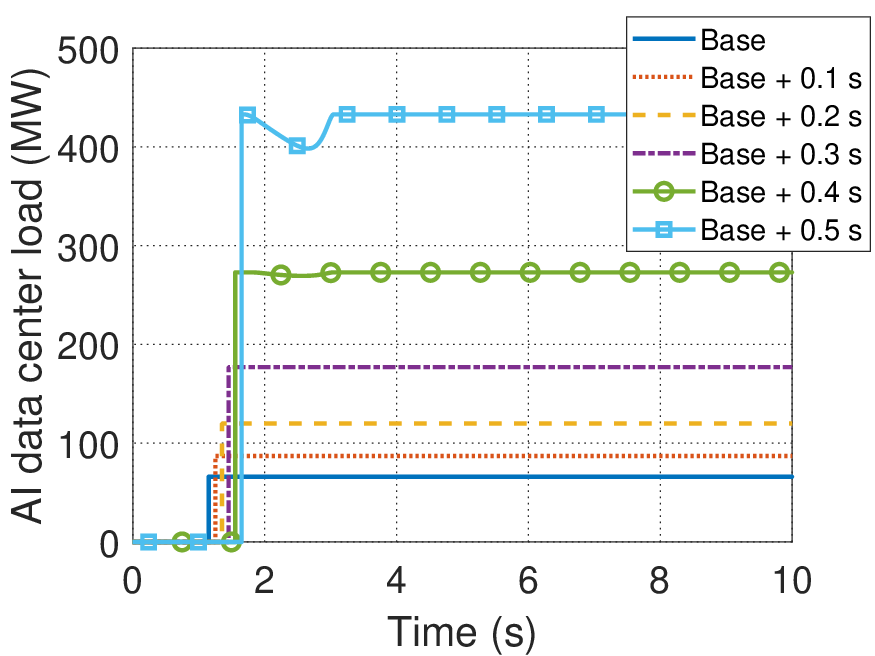}}
\caption{Effect of TILI activation delay on stabilizing the 715~MW
operating point in the SMIB system: (a) sending-end generator
rotor-angle responses and (b) TILI magnitude required to maintain
transient stability under the specified contingency.}\label{fig03}
\end{figure}

The same mechanism provides larger stability benefits as the available TILI response increases. Under the same SMIB conditions, 66 MW of TILI, corresponding to approximately 10\% of the 665 MW baseline generation limit, enables stable operation at 715 MW, as shown by the case labeled `Base' in Fig.~\ref{fig03}. This corresponds to an additional 50 MW of generation beyond the baseline stability-constrained limit.

Transmission reinforcement is a conventional measure for relieving transient-stability-constrained generation limits. For context, adding a third identical transmission line increases the modeled generation limit from 665 to 1,025~MW, while a 530~MW TILI response reaches the same limit under the specified contingency. This numerical comparison illustrates the potential magnitude of the stabilizing effect in this case and does not imply functional equivalence between TILI and transmission reinforcement.

The activation delay strongly affects the TILI magnitude required for stabilization. As shown in Fig.~\ref{fig03}, 66~MW of TILI is sufficient to maintain synchronism at the 715~MW operating point with the baseline activation delay of 0.05~s. The required magnitude increases nonlinearly to 87, 120, 177, 273, and 433~MW when additional delays of 0.1, 0.2, 0.3, 0.4, and 0.5~s are introduced, respectively. Thus, earlier activation during the post-fault first swing substantially reduces the load response required to stabilize the same operating condition.

\subsection{IEEE 39-Bus System}
To evaluate the applicability of TILI beyond the SMIB system, a modified IEEE 39-bus system \cite{ieee39bus} is configured to produce a transient-stability-constrained operating condition. A three-phase fault is applied at Bus~10 and cleared by tripping the transmission line between Buses~10 and~11. Under this contingency, the generator at Bus~32 exhibits the dominant post-fault rotor-angle excursion and is identified as the critical generator. Without TILI, increasing its output beyond 690~MW causes the post-fault rotor angle to diverge, resulting in loss of synchronism.

\begin{figure}[t]
\centering
\subfigure[One-line diagram and TILI locations]{
    \includegraphics[width=2.75in]{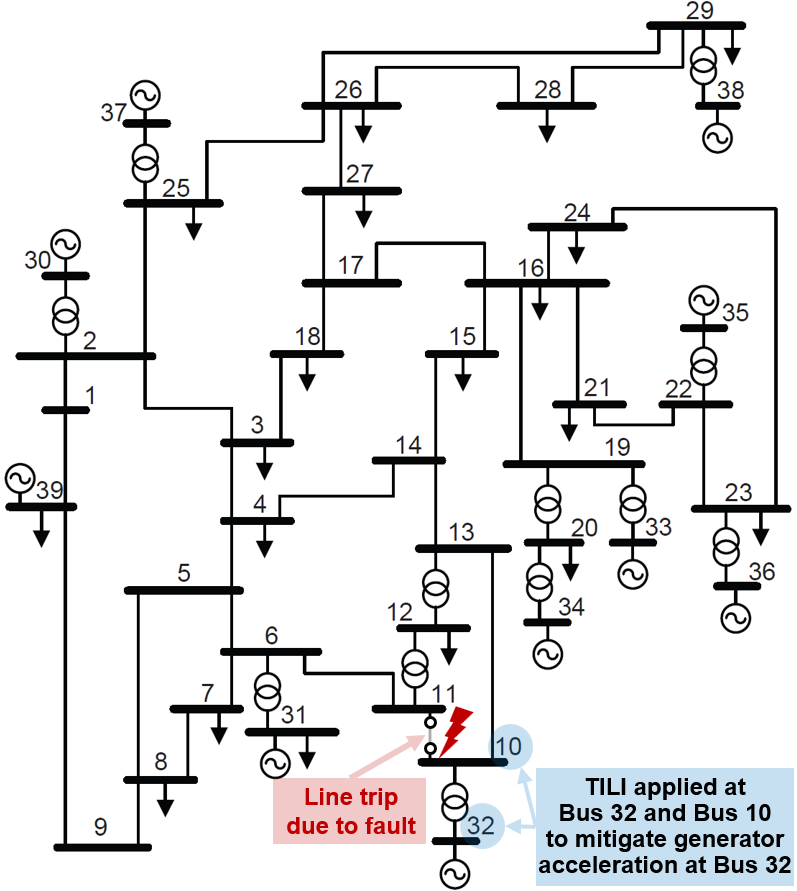}}
\subfigure[Increase in allowable generation at Bus 32]{
    \includegraphics[width=2.75in]{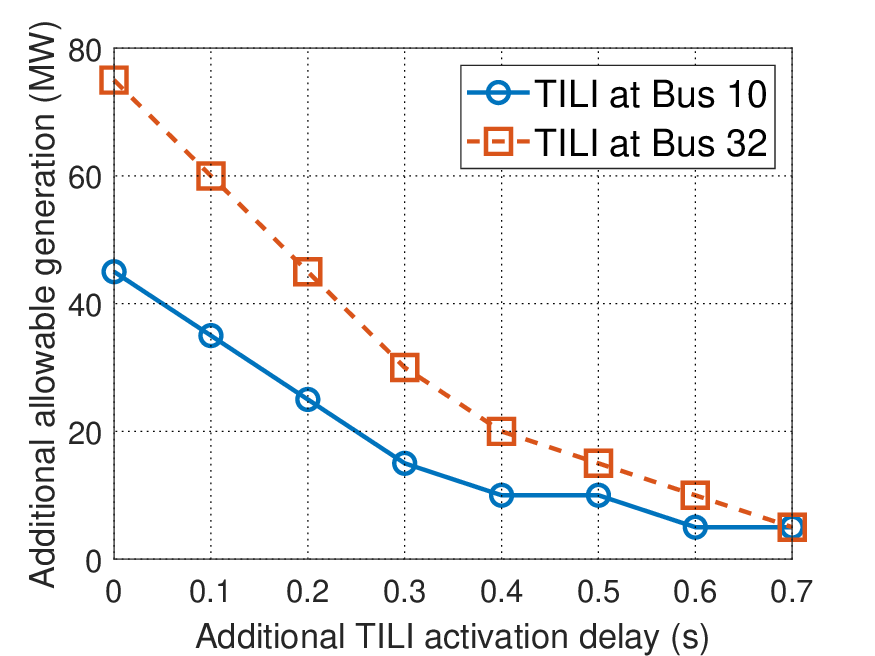}}
\caption{Effect of TILI location and activation delay on allowable
generation at Bus 32 in the IEEE 39-bus system: (a) one-line diagram
showing the specified contingency and TILI locations and (b)
generation-limit increase.}
\label{fig03_1}
\end{figure}

Fig.~\ref{fig03_1}(b) compares the increase in the stability-constrained output of the generator at Bus~32 when a 100~MW TILI response is applied separately at Bus~32 and Bus~10 under different activation delays. Both siting cases result in a higher stability-constrained output of the generator at Bus~32 than the no-TILI case. However, TILI applied at Bus~32 provides a larger generation-limit increase than the same response applied at Bus~10, demonstrating the importance of electrical siting. The Bus~10 case also improves stability, indicating that TILI can provide first-swing support even when it is not located at the terminal bus of the critical generator, although with reduced effectiveness.

The stabilizing effect decreases nonlinearly as the activation delay increases, while the performance difference between the two locations narrows at longer delays. These results suggest that TILI is most effective when deployed at an electrically effective bus and activated early after fault clearing. Thus, electrical siting and response timing are both important design considerations for using flexible AI data-center demand as a transient-stability resource.

\subsection{Korean Power System}
\begin{figure}[!t]
\centering
\subfigure[Korean transmission system highlighting the eastern generation and metropolitan load areas]{
    \includegraphics[width=2.35in]{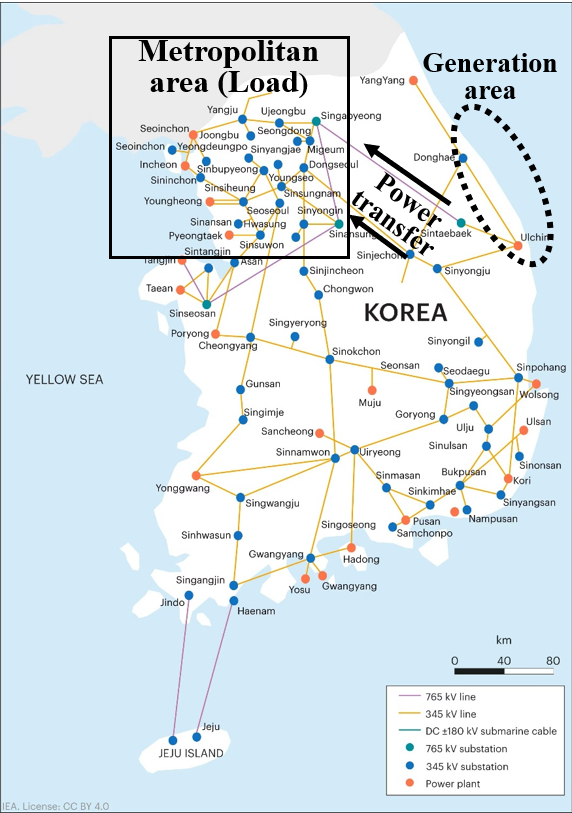}}
\subfigure[Systemwide rotor-angle spread with and without TILI]{
    \includegraphics[width=2.75in]{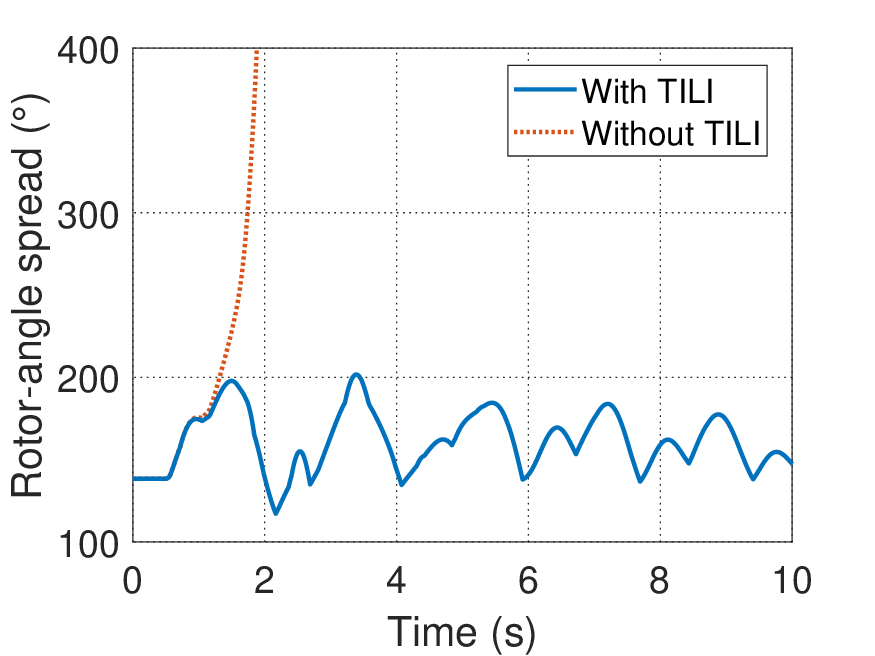}}
\caption{Korean transmission system and post-fault rotor-angle
responses under the specified 765-kV contingency: (a) transmission
network \cite{IEA_2020} highlighting the metropolitan load area and
eastern generation area and (b) systemwide rotor-angle spread with and without TILI.}
\label{fig04}
\end{figure}

The proposed TILI strategy is further evaluated using a large-scale Korean power system under operating conditions in which transient-stability constraints already limit interregional power transfer
\cite{KPS_ESS}.
The corresponding transmission network is shown in Fig.~\ref{fig04}(a)
\cite{IEA_2020}.
Electricity demand in Korea is highly concentrated in the metropolitan area, whereas a substantial share of nuclear and thermal generation is located along the eastern coast. Consequently, the metropolitan area relies heavily on power supplied from other regions because its local generation capacity is insufficient to meet regional demand \cite{Korea_status1,Korea_status2}. The eastern generation area considered in Fig.~\ref{fig04}(a) includes approximately 19.2~GW of installed nuclear and thermal generation capacity as of 2026, based on the unit-level generator data used in this study.

This regional imbalance requires large power transfers from the eastern generation area to the metropolitan load center. However, under a 765-kV transmission-line contingency considered in current Korean operational security assessments, some generators in the eastern region may lose synchronism, thereby limiting the allowable generation in the region
\cite{KPS_ESS}.
Consequently, aggregate generation in the eastern area is limited to approximately 11.5~GW, corresponding to about 60\% of its installed capacity. This case is used to evaluate whether TILI can increase the transient-stability-constrained generation limit in a large-scale transmission network.

\begin{figure}[!t]
\centering
\subfigure[Baseline delay]{\includegraphics[width=1.72in]{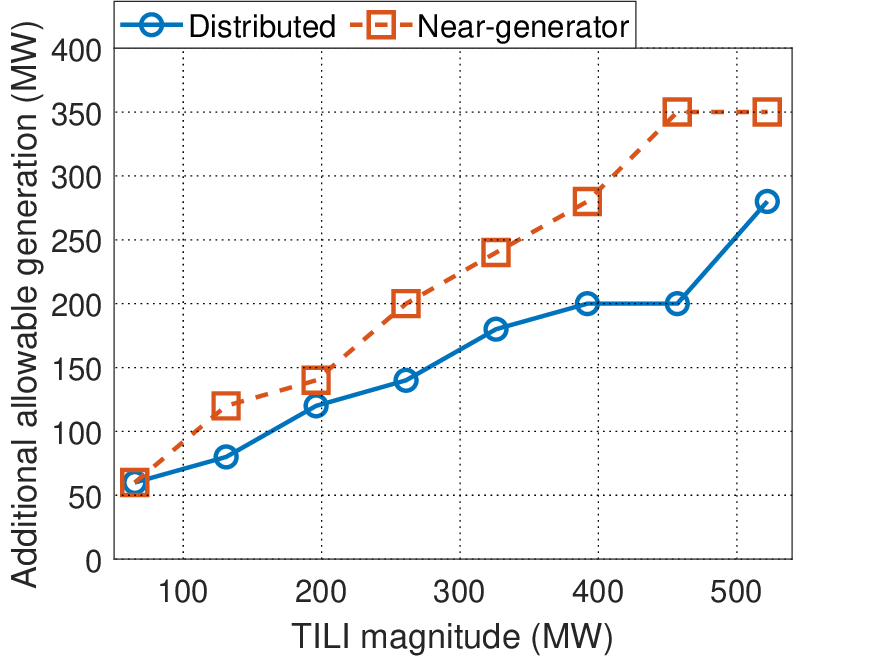}}
\subfigure[Baseline delay
$+\,0.2$~s]{\includegraphics[width=1.72in]{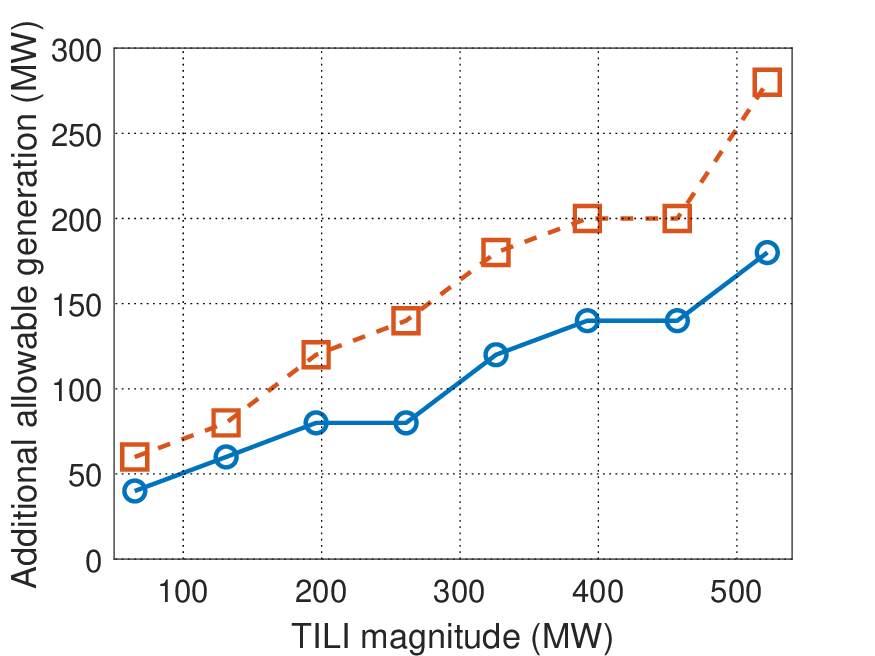}}
\subfigure[Baseline delay
$+\,0.4$~s]{\includegraphics[width=1.72in]{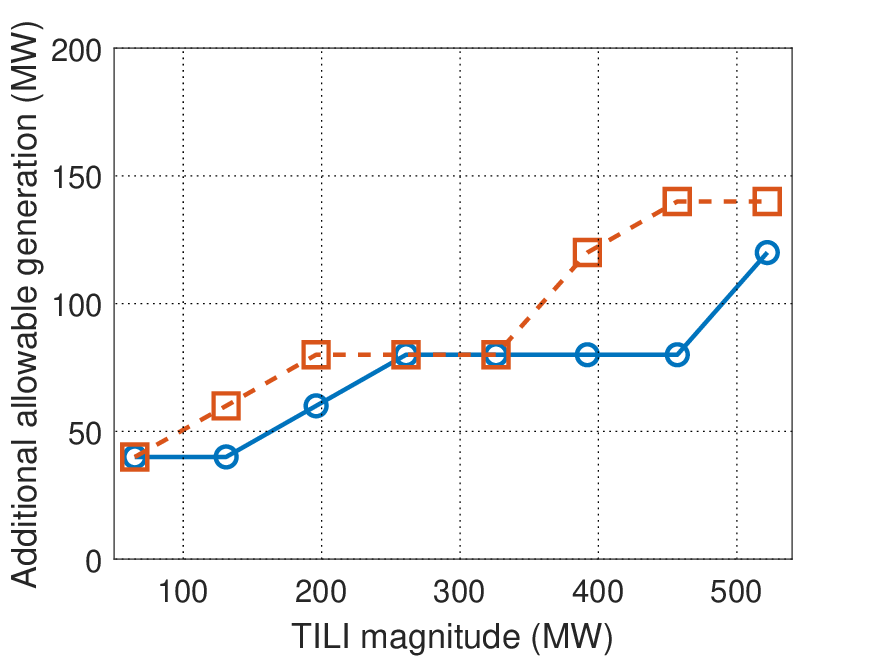}}
\subfigure[Baseline delay
$+\,0.6$~s]{\includegraphics[width=1.72in]{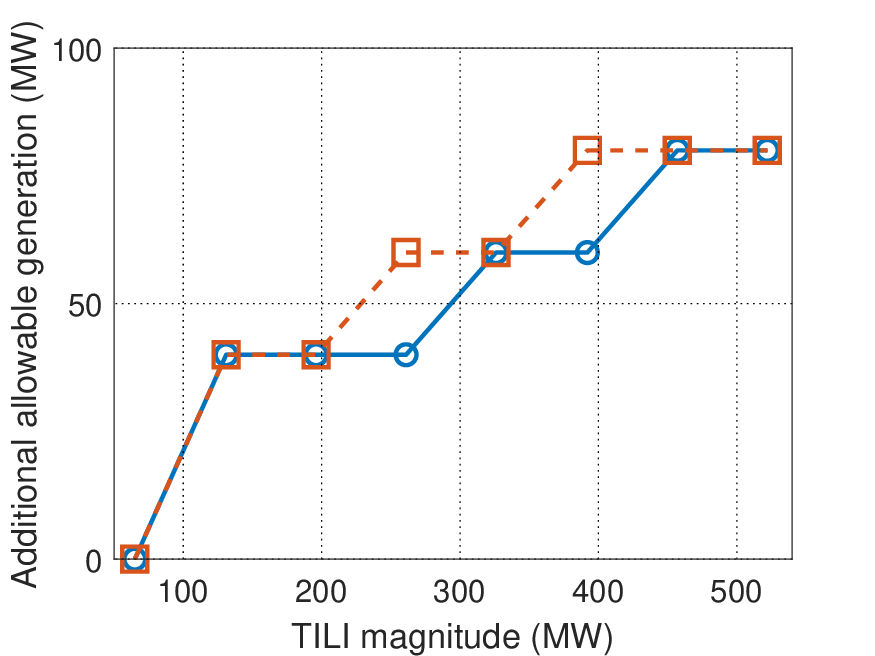}}
\caption{Effects of TILI response magnitude, electrical siting, and activation delay on the eastern-area generation limit in the Korean power system. The increase in the transient-stability-constrained generation limit is shown as a function of TILI magnitude under four activation-delay conditions: (a) baseline delay, (b) baseline delay
$+\,0.2$~s, (c) baseline delay $+\,0.4$~s, and (d) baseline delay $+\,0.6$~s. Note that the y-axis ranges differ across panels, but a consistent y-axis tick interval is maintained for comparison.}
\label{fig05}
\end{figure}

Fig.~\ref{fig04}(b) demonstrates the stabilizing effect of TILI through the systemwide rotor-angle spread. Under the same operating condition and contingency, the rotor-angle spread diverges without TILI, indicating loss of synchronism, whereas it remains bounded when TILI is activated. Beyond this qualitative comparison, Fig.~\ref{fig05} quantifies the extent to which TILI relaxes the eastern-area generation constraint as a function of TILI magnitude, activation delay, and electrical siting. Because large-scale AI data centers may not always be sited electrically close to the generators that dominate the post-fault instability, the distributed and near-generator configurations are compared.

Fig.~\ref{fig05}(a) shows that increasing the TILI magnitude at the baseline activation delay increases the eastern-area generation limit. For a given TILI magnitude, the near-generator configuration generally provides a larger generation-limit increase than the distributed configuration, demonstrating the importance of electrical siting. 

The stabilizing effect of TILI decreases nonlinearly as the activation delay increases, requiring a larger TILI magnitude to achieve the same generation-limit increase. Moreover, the performance difference between the two siting configurations narrows at longer delays. These results suggest that both electrical siting and activation timing are important, with the locational advantage being most pronounced when TILI is activated sufficiently early after fault clearing.

\begin{table*}[!t]
\caption{Comparison of Conventional Measures for Transient Stability Enhancement and TILI}
\label{tb01}
\centering
\footnotesize
\setlength{\tabcolsep}{3.5pt}
\renewcommand{\arraystretch}{1.15}

\begingroup
\renewcommand{\tabularxcolumn}[1]{m{#1}}

\begin{tabularx}{\textwidth}{@{}
>{\hsize=.65\hsize\centering\arraybackslash}X |
>{\hsize=1.18\hsize\raggedright\arraybackslash}X |
>{\hsize=.90\hsize\raggedright\arraybackslash}X |
>{\hsize=1.27\hsize\raggedright\arraybackslash}X
@{}}

\hline

\multicolumn{1}{c|}{\multirow{2}{*}{\textbf{Option}}} &
\multicolumn{1}{c|}{\multirow{2}{*}{\textbf{Main mechanism}}} &
\multicolumn{1}{c|}{\textbf{Activation or}} &
\multicolumn{1}{c}{\multirow{2}{*}{\textbf{Implementation constraints}}} \\

& &
\multicolumn{1}{c|}{\textbf{deployment timescale}} &
\\

\hline

Dynamic braking resistor
& Absorbs electrical power after a disturbance, thereby reducing generator acceleration.
& Approximately 100 ms.
& Requires dedicated equipment and additional capital investment. \\
\hline

ESS charging
& Increases electrical demand through battery charging, thereby absorbing electrical power and reducing accelerating power.
& Several hundred ms.
& Requires storage and inverter capacity; costs can be substantial at the hundreds-of-megawatts scale; sufficient charging headroom must be available. \\
\hline

Generator tripping
& Trips preselected generators after severe contingencies, reducing accelerating generation and limiting rotor-angle separation.
& Approximately 100 ms.
& Reduces available generation and may introduce frequency-stability risks if the generation loss is large. \\
\hline

Transmission reinforcement
& Increases transfer capability through new lines, upgraded conductors, transformers, or network expansion.
& Years to decades.
& Requires major capital investment, long lead times, permitting, and public acceptance. \\
\hline

TILI
& Initiates or resumes controllable AI training workloads near accelerating generators, increasing local electrical demand and generator electrical output.
& Potentially subsecond with pre-staged workloads; end-to-end validation is required.
& Requires flexible workload headroom, grid-triggered control, and validation of rapid load activation. \\
\hline

\end{tabularx}
\endgroup
\end{table*}

\section{Discussion}\label{discussion}
\subsection{Implications for Power System Stability Support}
Unlike conventional demand-response applications that primarily shift, smooth, or curtail demand over system-balancing timescales
\cite{ai_grid_asset,ai_power_stab}, the proposed TILI exploits upward demand flexibility during the critical post-fault period by initiating flexible training workloads, thereby reducing first-swing generator acceleration. Across the SMIB, IEEE 39-bus, and Korean power systems, the results consistently show that TILI is most effective when sufficient upward load-response capability is available at electrically effective locations and activated rapidly following a disturbance.

Importantly, TILI should be regarded as a complement to, rather than a replacement for, existing stability-enhancing measures. As summarized in Table~\ref{tb01}, braking resistors absorb excess electrical energy during disturbances
\cite{Braking_res},
ESSs operating in charging mode can provide a similar power-absorption response
\cite{ESS_swing},
and SPSs improve stability through corrective actions such as generator tripping, turbine-valve control, and controlled system separation
\cite{sps_all,sps_korea, EVA, seperation}.
TILI provides an additional corrective option by rapidly increasing controllable demand, thereby allowing accelerating generators to supply
more electrical power during the early post-fault period.

A distinguishing feature of TILI is that it uses the upward load flexibility of AI data centers installed primarily for computing services to support power system stability, rather than relying on a dedicated stability-support device. As large-scale AI data-center deployment is expected to expand, reserving a portion of their upward load-response capability for coordinated contingency response could enable these facilities to provide complementary power-system stability support. The proposed TILI may therefore complement transmission reinforcement and be coordinated with established corrective controls when verified upward load-response capability is available.

The Korean-system results further illustrate the potential value of this additional option. In the eastern generation region, generation remains constrained by transient stability despite the use of SPS-based generator-tripping
\cite{sps_korea}.
TILI introduces a demand-side corrective action that can supplement such existing measures and increase the generation that can remain online following a severe contingency.


\subsection{Deployment Requirements, Validation, and Applicability}
The results identify three principal requirements for practical TILI deployment. First, fast activation is essential because the stabilizing effect decreases rapidly as the activation delay increases. Delayed activation therefore requires a larger TILI response to provide the same stabilizing effect. Second, sufficient upward load-response capability must be available before a contingency occurs. This requires idle, paused, or pre-staged training jobs that can be rapidly initiated or resumed. Third, TILI siting should reflect the electrical influence of the additional demand on the generators that dominate the post-fault instability. Electrically remote loads may still provide stability support, but generally with reduced effectiveness.

Practical implementation requires coordination between power system operators and participating AI data centers. System operators need visibility into the committed TILI magnitude and its availability status. Participating data centers require control interfaces capable of receiving contingency-triggered signals and initiating or resuming pre-staged training jobs within the required time window. The committed TILI capability must also reflect the available electrical headroom and the operating status of flexible workloads. This coordination is analogous to the operational logic of SPSs \cite{sps_all}, but uses controllable demand as the corrective resource. 

The idealized step increase in active-power demand adopted in this study isolates the effects of response magnitude, activation delay, and electrical siting. Actual AI data-center responses may involve finite ramp rates, communication and scheduling delays, and sequential workload activation. Although large subsecond power variations have been observed in AI training workloads \cite{NERC_DS,LLM_cloud},
these observations do not yet demonstrate reliable, grid-triggered TILI activation at the multi-megawatt scale.

Representative component timescales relevant to TILI activation are estimated based on prior studies and technical documentation
\cite{cuda_graph,nvidia_800vdc,easyrider,ai_power_stab}.
For pre-staged GPU activation, these timescales correspond to an estimated aggregate response time of approximately 0.16 s after fault clearing, suggesting that subsecond TILI activation could be feasible. This estimate is close to the baseline delay plus 0.1 s (e.g., approximately 0.15 s) examined in the case studies, under which TILI retained substantial stabilizing effectiveness.
However, this estimate is component-based rather than an end-to-end demonstration. Future grid-triggered, multi-megawatt demonstrations should therefore validate the complete response chain, from disturbance detection and signal transmission to workload activation and verification of the delivered active-power response.

It should be highlighted that the stabilizing mechanism is not specific to AI data centers. Other large-scale flexible loads may provide similar transient-stability support if they can increase demand rapidly and reliably, maintain sufficient upward load-response capability, and are located at electrically effective buses. However, the benefits are expected primarily in systems where power transfer from a generation-rich area is limited by transient stability and the controllable load has a strong electrical influence on the accelerating generator group. Accordingly, system-specific dynamic studies are required before deploying TILI or analogous load-based corrective controls.

\subsection{Operational Suitability of AI Training Workloads for TILI}
From a power-system operational perspective, AI training workloads may be particularly suitable for the proposed TILI for two reasons: their operational flexibility and their ability to utilize available power headroom in mixed-use AI data centers. First, training workloads can provide greater temporal and scheduling flexibility than inference workloads. In general, AI inference workloads arise from real-time user requests and are therefore subject to strict latency and service-quality requirements
\cite{pipe}.
In contrast, AI model training is typically decoupled from real-time user requests, allowing adjustments in execution timing and scheduling
\cite{AI_DR,AI_scaler}. A recent study on demand response in AI data centers found that training workloads provided greater regulation flexibility than inference workloads, which was attributed to the longer and more malleable execution structure of training jobs
\cite{AI_DR}.
This capability has also been demonstrated through dynamic resource scaling and suspend--resume scheduling of real-world training jobs
\cite{AI_scaler}.

Second, AI training workloads can provide a practical means of utilizing available power headroom during inference operation without directly modulating latency-sensitive inference services. Measurements from operational LLM clusters showed that inference clusters retained substantially greater power headroom than training clusters
\cite{LLM_cloud},
indicating that inference operation may leave considerable capacity for additional electrical load. Furthermore, a GPU time-sharing system demonstrated that unused GPU capacity can be utilized by training workloads while preserving the performance of the primary inference service
\cite{pipe}.

Therefore, the flexibility of AI training workloads and the available headroom for additional electrical load during inference operation support their use as a controllable TILI resource in mixed-use AI data centers. Moreover, the additional energy consumed by TILI can be used to perform flexible training computations that would otherwise be scheduled later. During normal data-center operation, AI inference workloads can remain the primary service, and pre-staged training workloads can be activated when grid support is required without directly affecting inference service quality.

\section{Conclusion}\label{conclusion}
This paper proposed and evaluated TILI as a fast demand-side operating concept for transient-stability support in transmission-constrained power systems. Following fault clearing, TILI initiates or resumes flexible AI training workloads to rapidly increase active-power demand at AI data centers located at electrically effective buses. The resulting load increase allows accelerating generators to supply additional electrical power, thereby reducing the accelerating-power imbalance and limiting the first-swing rotor-angle excursion.

The SMIB system was first used to clarify the underlying mechanism of TILI. Its effectiveness was then evaluated in the IEEE 39-bus system and the large-scale Korean power system. Across all three systems, TILI increased the transient-stability-constrained generation limit under the studied contingencies. The extent of this increase depended on response magnitude, activation delay, and electrical siting. Earlier activation and siting at buses with a strong electrical influence on the critical generators consistently resulted in larger increases in the stability-constrained generation limit.

As large-scale AI data-center capacity is expected to expand,a portion of the associated upward load-response capability could potentially be coordinated as a complementary power-system stability resource. It should be noted that TILI is not intended to replace transmission reinforcement or established corrective controls; rather, it provides an additional demand-side option when sufficient electrical headroom, flexible workloads, and reliable grid-triggered activation are available.
Practical deployment will require system-specific dynamic studies and end-to-end validation of achievable TILI responses at large-scale AI data centers.

\ifCLASSOPTIONcaptionsoff
  \newpage
\fi

\bibliographystyle{IEEEtran}
\bibliography{IEEEabrv,TILI_JK}

\end{document}